\documentclass[10pt]{iopart}
\usepackage{iopams}	% provides some math fonts
\usepackage[english]{babel}
\usepackage{hyperref}
\usepackage{grffile}	% filenames can contain periods
\usepackage{graphicx}
\usepackage{cite}
\usepackage{multirow}	% in tabular
\usepackage{siunitx}	% formatting numerical values with \num

\usepackage{tikz}
\DeclareRobustCommand*\circled[1]{\tikz[baseline=(char.base)]{
            \node[shape=circle,draw,inner sep=.6pt] (char) {#1};}}

\newcommand*\euler{\mathrm{e}}

\newcommand\rmidx[2]{{#1}_{\mathrm{#2}}}

\newcommand*\diff{\mathrm d}
\newcommand*\funcdiff{\delta}

\newcommand\IU{\mathrm{i}}	% better use siunitx's \num{i}
\newcommand\conj[1]{{#1}^*}
\newcommand\cnormsq[2][2]{\left\vert{#2}\right\vert^{#1}}	% use as \cnormsq[]{c} to get the norm
\newcommand\cnorm[1]{\cnormsq[]{#1}}

\newcommand\LandauO[1]{\mathcal{O}\dep{#1}}
\newcommand\Fourier[1]{\mathfrak{F} #1}
\newcommand\dep[1]{\mathopen{}\left( #1 \right)\mathclose{}}
\newcommand\funcdep[1]{\mathopen{}\left[ #1 \right]\mathclose{}}

\let\oldvec\vec
\renewcommand\vec[1]{\oldvec{\mathbf{#1}}}

\newcommand*\vecx{\vec{x}}
\newcommand*\veck{\smash{\vec{k}}\vphantom{\vec{x}}}
\newcommand*\dirint{\int \diff^2 x \,}

\newcommand*\den{\psi}
\newcommand*\ori{U}
\newcommand*\freeen{\mathcal{F}}
\newcommand*\modorifunc{f}
\newcommand*\couplingop{\kappa}
\newcommand*\paramBl{\rmidx{B}{l}}
\newcommand*\paramBx{\rmidx{B}{x}}
\newcommand*\paramD{D}
\newcommand*\paramE{E}
\newcommand*\paramF{F}
\newcommand*\timestep{\Delta t}

\newcommand\allparam[6]{$n=#1, \paramBl=\num{#2}, \paramBx=\num{#3}, \paramD=\num{#4}, \paramE=\num{#5}, \paramF=\num{#6}$}
\usepackage{xspace}

\renewcommand*\etal{\emph{et\,al.}\@\xspace}
\newcommand*\ie{\emph{i.\,e.}\@\xspace}
\newcommand*\eg{\emph{e.\,g.,}\@\xspace}

\newcommand*\PFC{PFC\xspace}
\newcommand\fold[1]{\ensuremath{#1}-fold\xspace}
\usepackage{xcolor}
\begin{document}

\title%
	[PFC for \fold{n} particles]%
	{Phase Field Crystal model for particles with \fold{n} rotational symmetry in two dimensions}
\author{%
	Robert F. B. Weigel and
	Michael Schmiedeberg
}
\address{%
	Institut für Theoretische Physik,
	Friedrich-Alexander-Universität Erlangen-Nürnberg,
	Staudtstr. 7, 91058 Erlangen, Germany
}
\ead{\mailto{robert.rw.weigel@fau.de}, \mailto{michael.schmiedeberg@fau.de}}
\begin{abstract}
	We introduce a Phase Field Crystal (PFC) model for particles with $n$-fold rotational symmetry in two dimensions. 
	Our approach is based on a free energy functional that depends on the reduced one-particle density,
	the strength of the orientation, and the direction of the orientation, where all these order parameters depend on the position.
	The functional is constructed such that for particles with axial symmetry (\ie $n=2$) 
	the PFC model for liquid crystals as introduced 
	by H. L\"owen [\emph{J. Phys.: Condens. Matter} \textbf{22}, 364105 (2010)] is recovered.
	We discuss the stability of the functional and explore phases that occur for $1\leq n\leq 6$. 
	In addition to isotropic, nematic, stripe, and triangular order, 
	we also observe cluster crystals with square, rhombic, honeycomb, and even quasicrystalline symmetry.
	The $n$-fold symmetry of the particles corresponds to the one 
	that can be realized for colloids with symmetrically arranged patches.
	We explain how both, repulsive as well as attractive patches, are described in our model.
\end{abstract}

%\keywords{}	% Macro does not work!
\begingroup
	\vspace{28pt plus 10pt minus 18pt}
	\par
\endgroup

\maketitle
%\ioptwocol	% two-column text

%textwidth in cm: \printinunitsof{cm}\prntlen{\textwidth}
%textheight in cm: \printinunitsof{cm}\prntlen{\textheight}

%\input{introduction.ltx}
\section{Introduction}

The formation of patterns in particulate systems is a long-standing topic \cite{cross1993}. 
A mean field approach to describe the formation of complex equilibrium phases is the so-called Phase Field Crystal (\PFC) model\cite{elder2002,elder2004}
where a free energy expansion in a density-like field and its gradient is considered in a way similar as 
in the well-known approaches by Swift and Hohenberg \cite{swift1977}, by Alexander and McTague \cite{alexander1978}, 
or by Lifshitz and Petrich \cite{Lifshitz1997}.
Non-trivial phases are stabilized in \PFC models in various ways, \eg by using more than one length scale \cite{Achim2014,Ratliff2019} 
(similar as in the Lifshitz-Petrich model \cite{Lifshitz1997,Barkan2011,Barkan2014}) 
or by introducing a competition with an incommensurate external potential \cite{Schmiedeberg2006,Rottler2012}.
In another approach anisotropic particles are considered.
These would break the flow of the text, as they do not introduce an orientational field.
Instead they directly couple the director field to the density field.
By introducing an orientational field 
and couplings between the orientational and the density-like field, \PFC models for particles with 
polar \cite{WittkowskiPRE2011} or axial \cite{LoewenJPCM2010,AchimPRE2011,CremerEPL2012} symmetry have been modeled.
Here we want to generalize these models for particles with other $n$-fold rotational symmetries.
Note, there are other \PFC approaches to describe systems composed of particles
with a certain rotational symmetry \cite{WangPRE2018,MkhontaEPL2013}.
However, in these approaches the orientation field and the density field cannot vary independently and as a consequence 
some phases like plastic crystals and oriented crystals cannot be distinguished.

Colloids with a given rotational symmetry can be realized by decorating the particles with attractive or repulsive patches. 
These so-called patchy colloids are known to exhibit a complex phase behavior \cite{doye2007,glotzer2007,pawar2010,doppelbauer2010,bianchi2011,doppelbauer2012}.
Patchy colloids can even be used to obtain quasicrystals \cite{vanderlinden2012,reinhardt2013,GemeinhardtEPJE2018,GemeinhardtEPL2019} 
or to be designed in a way to obtain a complex ordering as desired \cite{Tracey2019,romano2020,Tracey2021,noya2021}.

\begin{figure}
	\centering
	\includegraphics{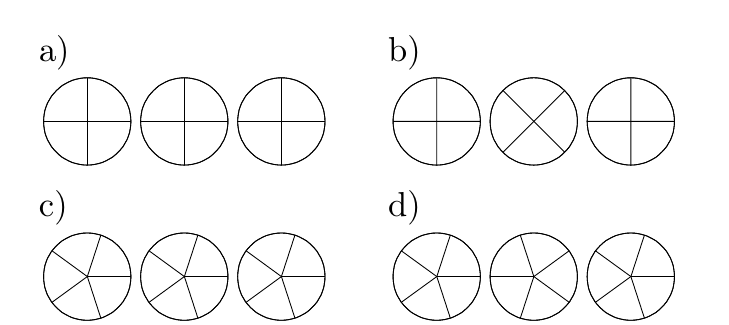}
	\caption{
		Sketch of the preferred orientation of neighboring particles with (a,b) $n=4$-fold rotational symmetry 
		showing cases that are typical for all even $n$ 
		and (c,d) $n=5$-fold rotational symmetry, which is typical for odd $n$.
		In (a,c) neighboring particles prefer to possess the same orientation, while in (b,d) they prefer alternating orientations, 
		which we realize by a modulated alignment interaction. 
		As a consequence, the cases in (a,d) correspond to patchy particles with attractive patches, while in (b,c) the patches are repulsive.
	}
	\label{F:Schematic}
\end{figure}

In this work we want to consider both, attractive and repulsive patches.
In case of attractive patches the patches of neighboring particles tend to point towards each other
while repulsive patches tend to be oriented away from neighboring patches.
As a result neighboring particles either have the same orientation 
or they are rotated by an angle $\pi/n$ as illustrated in \fref{F:Schematic}.

Note that the particles in this work do not possess any hard core. Therefore, large overlaps can occur 
and the ordering that we report corresponds to those 
of so-called cluster crystals \cite{Sciortino2004,Mladek2006,Lenz2012,Sciortino2013}.
Cluster crystals occur naturally in \PFC models or related approaches \cite{Barkan2014} 
and can form periodic as well as aperiodic structures \cite{Barkan2014}. 
In the conclusions in \sref{S:Conclusions} we will discuss how our approach might be modified 
in order to describe particles with hard cores, \ie particles that cannot overlap significantly.

The article is organized as follows: In \sref{S:Model} we introduce the \PFC model for $n$-fold particles 
and explain how we determine the stable phases numerically.
In \sref{S:Results} we first present an overview of the phases that occur for various rotational symmetries 
before we discuss these phases in more detail.
Finally, we conclude in \sref{S:Conclusions}.
%%%%%%%%%%%%%%%%%%

%\input{model.ltx}
\section{Model}
\label{S:Model}

We generalize the \PFC models that have been introduced for particles with 
\fold{1} \cite{WittkowskiPRE2011} or \fold{2} \cite{LoewenJPCM2010,AchimPRE2011,CremerEPL2012} symmetry
to particles with \fold{n} symmetry. We will consider both, attractive and repulsive patches. 
Concerning the notation, we will follow the model presented in \cite{AchimPRE2011}, which is shortly outlined in the next subsection.

\subsection{Short summary of the \PFC model for \fold{2} symmetry}
\label{S:ModelAchim}
%% free energy of Achim \etal
In \cite{AchimPRE2011} Achim \etal have proposed and studied a \PFC model for apolar liquid crystals,
\ie particles of \fold{2} rotational symmetry.
The used free energy functional model had previously been derived by Löwen in \cite{LoewenJPCM2010} from classical density functional theory.
The free energy $\freeen \funcdep{\den, \ori}$ is given as a functional of the density-like field $\den$ and the orientation field $\ori$.
%These fields are an expansion of the 1-particle density.
The density-like field $\den$ gives the deviation from the mean density and
the orientational field is defined in terms of the nematic order parameter $\psi_2$ and nematic director field $\varphi$ as
\begin{equation}
	\ori\dep{\vecx} = \psi_2\dep{\vecx} \exp\dep{\IU 2\varphi\dep{\vecx}}.
	\label{E:DefineOri}
\end{equation}
Hence the modulus of $\ori$ represents the intensity of orientational ordering, while the complex phase encodes the direction.
In slightly modified notation, the free energy is
\begin{equation}
	\eqalign{%
		\freeen \funcdep{\den, \ori} =
		\dirint \Bigg(
		& \paramBl \den^2 + \paramBx \den \left( 2\nabla^2 + \nabla^4 \right) \den - \frac{1}{3} \den^3 + \frac{1}{6} \den^4 \cr
		& + \paramD \cnormsq{\ori} - \paramE \Re\dep{ \ori \nabla^2 \conj{\ori} } + \frac{1}{256} \cnormsq[4]{\ori} \cr
		& + \paramF \Re\dep{\ori \couplingop^2} \den + \frac{1}{8}\left(\den -1\right)\den\cnormsq{\ori}
		\Bigg)
	}
	\label{E:FreeEnergyAchim}
\end{equation}
with the complex derivative operator
\begin{equation}
	\couplingop = \partial_x - \IU \partial_y.
\end{equation}
The values of the parameters $\paramBl$ through $\paramF$ can be linked to the temperature, mean density, and the interaction of the modeled particles.
In the scope of \PFC modeling they can be considered free parameters.
%% minimization
The equilibrium phases are found by minimizing the free energy with respect to the fields $\den$ and $\ori$.
In the minimization, the mean of the density-like field is conserved, whereas the orientation field is treated as a non-conserved field.
Therefore the pseudodynamical equations read
\begin{eqnarray}
	\frac{\partial \den}{\partial t} = - \frac{\funcdiff \freeen}{\funcdiff \den} + \lambda\dep{t} \\
	\frac{\partial \ori}{\partial t} = - \frac{\funcdiff \freeen}{\funcdiff \conj{\ori}}
\end{eqnarray}
with the Lagrange multiplier $\lambda\dep{t}$, keeping the mean of $\den$ constant.
%Instead of the Lagrange multiplier, $\frac{\partial \den}{\partial t} = \nabla^2 \frac{\funcdiff \freeen}{\funcdiff \den}$ would have the same effect.
%But the treatment of the non-linear terms would be more complicated.

\subsection{New generalization for \fold{n} symmetry}
We now want to generalize the free energy functional of \eref{E:FreeEnergyAchim} to \fold{n} symmetry. 
First, the definition of the orientation field from \eref{E:DefineOri} is generalized to
\begin{equation}
	\ori\dep{\vecx} = \psi_n\dep{\vecx} \exp\dep{\IU n\varphi\dep{\vecx}}.
	\label{E:RedefineOri}
\end{equation}
In this way the interpretation of $\ori$ is compatible with \fold{n} symmetry, since the complex phase of $\ori$ is $2\pi/n$-periodic in the director field $\varphi$.
This means $\ori$ performs a full turn in the complex plane when the particle orientation rotates by $2\pi/n$.

Note that the term proportional to $\paramF$ ($\paramF$-term) in \eref{E:FreeEnergyAchim} is the only contribution to the free energy 
that depends on the complex phase of $\ori$.
Since all terms in $\freeen\funcdep{\den, \ori}$ need to be invariant under local rotations of the coordinate system,
the $\paramF$-term needs to be adjusted to the new definition of $\ori$.
The operator $\couplingop$ transforms under a local rotation of the coordinate system by an angle $\alpha$ as $\couplingop' = \couplingop \euler^{-\IU \alpha}$.
Therefore, $\ori \couplingop^n$ is the simplest term that is linear in $\ori$ and that respects local \fold{n} rotational symmetry.
Hence the $\paramF$-term of \eref{E:FreeEnergyAchim} is generalized to
\begin{equation}
	\paramF \Re\dep{\ori \couplingop^n} \den
\end{equation}

A modification of the $\paramE$-term proves to be necessary for numerical stability 
as we will explain in more detail in our discussion of the stability presented in \sref{S:Stability}.
Hence we replace the operator $-\nabla^2$ in the $\paramE$-term by $\modorifunc\dep{\nabla^2}$.
Moreover, this generalization permits modulated alignment that will be used to reflect both repulsive as well as attractive patches 
as will be explained in \sref{S:Alignment}.

With the $\paramE$-term and $\paramF$-term modified, the free energy becomes
\begin{equation}
	\eqalign{%
		\freeen \funcdep{\den, \ori} =
		\dirint \Bigg(
		& \paramBl \den^2 + \paramBx \den \left( 2\nabla^2 + \nabla^4 \right) \den - \frac{1}{3} \den^3 + \frac{1}{6} \den^4 \cr
		& + \paramD \cnormsq{\ori} + \paramE \Re\dep{ \ori \modorifunc\dep{\nabla^2} \conj{\ori} } + \frac{1}{256} \cnormsq[4]{\ori} \cr
		& + \paramF \Re\dep{\ori \couplingop^n} \den + \frac{1}{8}\left(\den -1\right)\den\cnormsq{\ori}
		\Bigg).
	}
	\label{E:FreeEnergyEmodFn}
\end{equation}
This is the new free energy functional for particles with \fold{n} rotational symmetry that we study in this article.

%% identify terms
%% also point out that more complicated terms can be added
When keeping $\modorifunc\dep{\nabla^2} = -\nabla^2$ and specializing to $n = 2$, our model reduces to the one from \cite{AchimPRE2011}.
Moreover, the generalization proposed here is in accordance with a model for
polar liquid crystals, \ie particles of \fold{1} rotational symmetry ($n = 1$),
presented by Wittkowski \etal in \cite{WittkowskiPRE2011}:
For example, the $\paramF$-term for $n = 1$ can be identified with the term proportional to $B_1$ in \cite{WittkowskiPRE2011}.
\Tref{T:CompareLiteratureTerms} provides a detailed comparison of all terms.
Note that although \cite{WittkowskiPRE2011} considers polar particles of \fold{1} symmetry,
contributions of \fold{2} (nematic) symmetry are taken into account in \cite{WittkowskiPRE2011},
by introducing fields for both, polar and nematic order parameter.
% the director fields are assumed to align.
This is reasonable, since the \fold{2} symmetry is compatible with the underlying \fold{1} symmetry.
The \fold{2} contributions can thus be seen as an extension upon the lowest-order orientational ordering.
Yet, the number of parameters is drastically increased due to the multitude of cross terms.
We narrow down the number of terms in the free energy functional by restriction to a single orientation field, 
representing the fundamental contribution of \fold{n} symmetry.

\begin{table}
	\caption{\label{T:CompareLiteratureTerms}%
		Comparison of terms occurring in the excess free energy for polar (\fold{1}) particles \cite{WittkowskiPRE2011}, including nematic (\fold{2}) contributions,
		with the respective terms for purely \fold{2} \cite{LoewenJPCM2010} and \fold{1} symmetry.
		The single terms are referenced by their coefficients in the notation of the respective publication.
		Terms that cannot occur in a model are marked with a dash (---) and deviations are specified.
		The terms of \cite{WittkowskiPRE2011}, proportional to $E_1$ through $G_7$, are unparalleled in \cite{LoewenJPCM2010} 
        and this work.
		Moreover the terms in \cite{AchimPRE2011} are identical to those in \cite{LoewenJPCM2010},
		up to a rescaling, which merges $B$ and $C$ of \cite{LoewenJPCM2010} into $\paramBx$ of \cite{AchimPRE2011}.
	}
	\begin{indented}
	\item[]%
	\begin{tabular}{ccc}
			\br
			$n=2$ \cite{LoewenJPCM2010} & $n=1 \& 2$ \cite{WittkowskiPRE2011} & $n=1$ [this work]\\
			\br
			$A$ & $A_1$ & $\paramBl$ \\
			$B$ & $A_2$ & \multirow{ 2}{*}{${\Big\rbrace} \paramBx$} \\
			$C$ & $A_3$ &  \\
			\mr
			--- & $B_1$ & $\paramF$, up to an irrelevant sign \\
			--- & $B_2$ & --- \\
			$F$, lacking contributions of $\nabla\varphi$ & $B_3$ & --- \\
			\mr
			--- & $C_1$ & $\paramD$ \\
			--- & $C_2$ & $\paramE$, up to modifications \\
			--- & $C_3$ & lacking \\
			\mr
			$D$ & $D_1$ & --- \\
			$E$ & $D_2$ & --- \\
			\br
		\end{tabular}
	\end{indented}
\end{table}

\subsection{Minimization scheme and numerical details}
Motivated by the numerical investigation of systems with axial symmetry by Achim \etal \cite{AchimPRE2011} 
we implement a similar combination of explicit Euler integration in direct space for the contributions from $\rmidx\freeen{dir}$ and
implicit Euler integration in reciprocal space for the contributions from $\rmidx\freeen{rec}$.
Conveniently, the terms of $\freeen\funcdep{\den, \ori}$, that are quadratic or bilinear in $\den$ or $\ori$, can be expressed in reciprocal space,
where spatial derivation reduces to multiplication with wave vector components.
%\begin{equation}
%	\eqalign{%
%		\rmidx\freeen{rec}\funcdep{\Fourier\den, \Fourier{\Re\ori}, \Fourier{\Im\ori}} = \recint \Bigg( 
%		& \bigg( \paramBl + \paramBx \left( -2 \veck^2 + \veck^4 \right) \bigg) \cnormsq{\Fourier\den} \\
%		& + \bigg( \paramD + \paramE \modorifunc\dep{-\veck^2} \bigg) \big( \cnormsq{\Fourier{\Re\ori}} + \cnormsq{\Fourier{\Im\ori}} \big) \\
%		& + \paramF \bigg( \conj{\Fourier{\Re\ori}} \Re\dep{\couplingop^n} - \conj{\Fourier{\Im\ori}} \Im\dep{\couplingop^n} \bigg) \Fourier\den
%		\Bigg)
%	}
%\end{equation}

To be specific, the explicit Euler integration in direct space is given by
\newcommand*\helpden{\den\prime}
\newcommand*\helpori{\ori\prime}
\begin{eqnarray}
	\helpden = 
	\den - \timestep \frac{\funcdiff \rmidx\freeen{dir}}{\funcdiff \den}\funcdep{\den, \Re\ori, \Im\ori}, \\
	\Re\helpori = 
	\Re\ori - \timestep \frac{\funcdiff \rmidx\freeen{dir}}{\funcdiff \Re\ori}\funcdep{\den, \Re\ori, \Im\ori}, \\
	\Im\helpori = 
	\Im\ori - \timestep \frac{\funcdiff \rmidx\freeen{dir}}{\funcdiff \Im\ori}\funcdep{\den, \Im\ori, \Im\ori}.
\end{eqnarray}
and the implicit Euler integration in reciprocal space leads to an evolution according to
\begin{equation}
	\pmatrix{ \Fourier{\den\prime\prime} \cr \Fourier{\Re\ori\prime\prime} \cr \Fourier{\Im\ori\prime\prime} } = 
	\left( 1 - \timestep L \right)^{-1}
	\pmatrix{ \Fourier{\helpden} \cr \Fourier{\Re\helpori} \cr \Fourier{\Im\helpori} }
	\label{E:ImplicitEuler}
\end{equation}
where $\Fourier$ denotes the Fourier transformation and 
$L$ is a matrix
%\begin{equation}
%	L = \pmatrix{
%	-2\left( \paramBl +\paramBx \left( -2\veck^2 + \veck^4 \right) \right) & -\paramF \conj{\Re\dep{\couplingop^n}} & -\paramF \conj{\Im\dep{\couplingop^n}} \cr
%	-\paramF \Re\dep{\couplingop^n} & -2\left( \paramD + \paramE \modorifunc\dep{-\veck^2} \right) & 0 \cr
%	-\paramF \Im\dep{\couplingop^n} & 0 & -2\left( \paramD + \paramE \modorifunc\dep{-\veck^2} \right)
%	}
%\end{equation}
that contains coefficients stemming from the functional derivatives of $\rmidx\freeen{rec}$.
The terms in the matrix $L$ will be shortly discussed in the stability analysis in section \ref{S:Stability}.
After the implicit integration, the Lagrange multiplier is applied,
setting the appropriate value for the $\veck = 0$-component of $\Fourier{\den\prime\prime}$.
The backwards transformation to direct space completes the timestep.

We apply periodic boundary conditions and discretize $\den$ and $\ori$ on a grid of \num{512 x 512} or \num{1024 x 1024} points.
The intrinsic lengthscale is set by the $\paramBx$-term to $k = 1$, \ie a peak-peak distance of $2\pi$ in direct space.
The system size spans 10 to 14 peak-peak distances,
such that single peaks, as well as structures much larger than the typical unit cell are resolved.
We perform several computations for each point in phase space, starting from random noise in both fields.
When the resulting free energies differ, which happens when defects or metastable structures appear,
we start more computations with appropriately patterned initial fields.
The structure that yields the lowest free energy is assumed to be the equilibrium phase.

\subsection{Modulated or unmodulated alignment to implement attractive or repulsive patches}
\label{S:Alignment}

Here we explain how the function $\modorifunc\dep{\nabla^2}$ (or $\modorifunc\dep{-k^2}$ in reciprocal space) 
is chosen such that either the alignment or the anti-alignment of neighboring particles is preferred. 
Note that the differences in alignment of neighboring particles is used to model repulsive or attractive patches 
as sketched and explained in Fig. \ref{F:Schematic}.
The asymptotics of $\modorifunc\dep{-k^2}$, \ie the behavior at large wave vectors, are dictated by the stability requirement 
which will be discussed in section \ref{S:Stability}.

In case neighboring particles prefer to align, we use a monotonic function namely
\begin{equation}
	\modorifunc\dep{-k^2} = k^m
\end{equation}
with $m = \max \left\lbrace {2n - 4, 2} \right\rbrace$. Note that for $n=2$ the term used in \cite{AchimPRE2011} is obtained.

In case the anti-alignment of neighboring particles is preferred, the function $\modorifunc\dep{-k^2}$ should be non-monotonic 
leading to a negative free energy contribution for an orientation that alternates on the length given by the nearest neighbor distance.
As anti-aligned particles experience a rotation of $\pi/n$ from one particle to its neighbor,
the period of $\ori$ amounts to 2 particle distances.
This corresponds to a wavelength of $k = \frac{1}{2}$.
However, we aim not to introduce a second lengthscale via the preferred wavelength of alignment.
Thus we choose an extended in $\modorifunc\dep{-k^2}$, covering $k = \frac{1}{2}$ and $k = 1$.
Furthermore, the asymptotics should be similar as for the alignment case.
Based on these requirements we employ a continuous piecewise-defined term namely
\begin{equation}
	\modorifunc\dep{-k^2} = \cases{
		-\left( k / \num{.4} \right)^2	& for \( k^2 \leq \num{.4}^2 \)	\\
		-1                            	& for \( \num{.4}^2 < k^2 \leq \num{1.1}^2 \) \\
		k^m -\num{1.1}^m -1           	& for \( \num{1.1}^2 < k^2 \).
	}
	\label{E:ModOri}
\end{equation}
We call this case the modulated alignment case.
Since $\modorifunc\dep{0} = 0$, the parameter $\paramD$ has the same meaning for modulated 
and unmodulated alignment and is not shifted by $\paramE$.
%%%%%%%%%%%%%%%%%%

%\input{results.ltx}
\section{Results}
\label{S:Results}

\subsection{Stability requirement}
\label{S:Stability}

To study the stability we have a closer look at the implicit integration step \eref{E:ImplicitEuler} 
that corresponds to the linear contribution of the integration and is governed by a matrix
%The matrix in the imlicit Euler integration step of \eref{E:ImplicitEuler} is
%the matrix is
%\begin{equation}
%	1 - \timestep L =
%	\pmatrix{
%	a & \conj{c_R} & -\conj{c_I} \cr
%	c_R & d & 0 \cr
%	-c_I & 0 & d
%	}
%\end{equation}
%and its inverse is
\begin{eqnarray}
	\fl %
	\left( 1 - \timestep L \right)^{-1} = \frac1{ad-\cnormsq{c_R}-\cnormsq{c_I}} \pmatrix{
		d & -\conj{c_R} & \conj{c_I} \cr
		-c_R & a - \cnormsq{c_I} /d & -c_R \conj{c_I} /d \cr
		c_I & -\conj{c_R} c_I /d & a - \cnormsq{c_R} /d
	}
	\label{E:MatrixElements} 
\end{eqnarray}
with
\begin{eqnarray}
	a = 1 + 2 \timestep \left( \paramBl +\paramBx \left( -2\veck^2 + \veck^4 \right) \right) \\
	c_R = \timestep \paramF \IU^n \Re\dep{\left( k_x - \num{i} k_y \right)^n} \\ %\Re\dep{\couplingop^n} \\
	c_I = - \timestep \paramF \IU^n \Im\dep{\left( k_x - \num{i} k_y \right)^n} \\ %\Im\dep{\couplingop^n} \\
	d = 1 + 2 \timestep \left( \paramD + \paramE \modorifunc\dep{-\veck^2} \right)
\end{eqnarray}
We now want to choose $\modorifunc\dep{\nabla^2}$ such that the integration is stable.
Note that in general the choice that is used in \cite{LoewenJPCM2010,AchimPRE2011} is not suitable:
For example, if $n \geq 4$ and the $\paramE$-term was given by $\modorifunc\dep{\nabla^2} = -\nabla^2$,
the computations would exhibit a numerical instability for many (significant) regions in physical parameter space.

\begin{figure}
	\centering
	\includegraphics{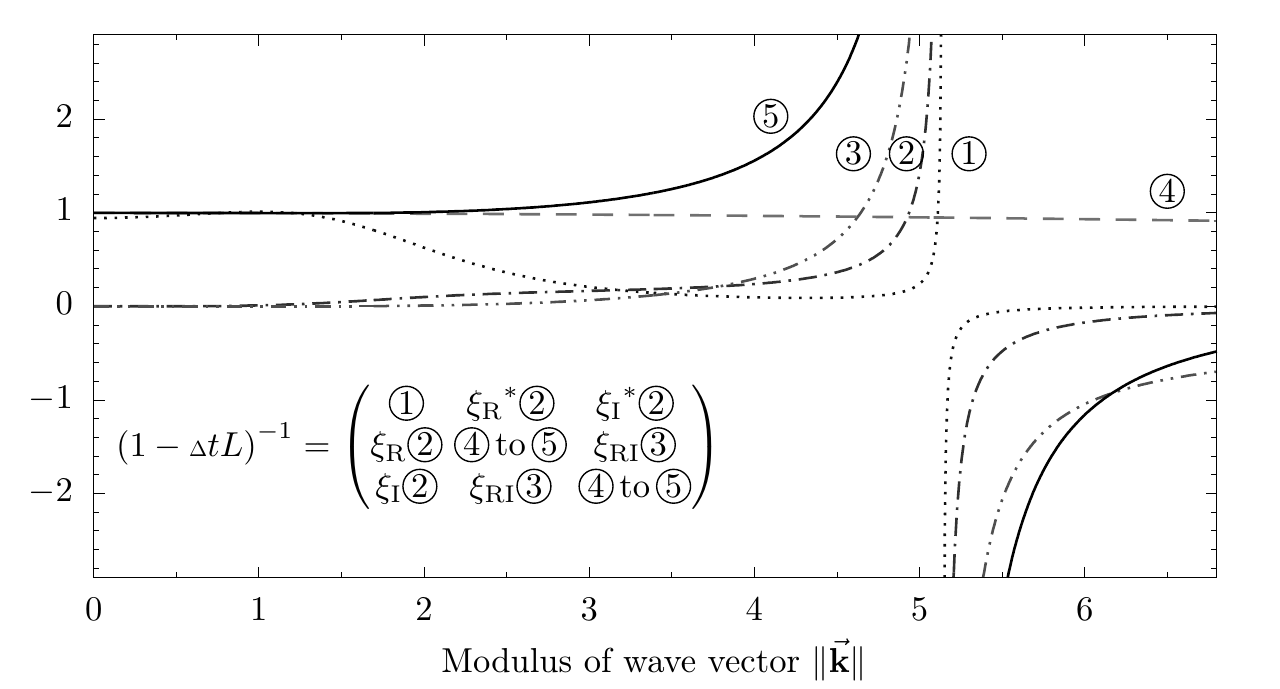}
	\caption{
		Matrix elements from \eref{E:MatrixElements} as functions of the wave vector.
		Plotted are the most extreme contours, which are observed along certain radial directions in reciprocal space.
		The diagonal elements vary between functions \circled{4} and \circled{5}, depending on direction.
		The factors $\rmidx\xi{R}$, $\rmidx\xi{I}$, $\rmidx\xi{RI}$ arise from the real and imaginary part of $\couplingop^n$.
		They are $2\pi/n$-periodic in the polar angle associated with direction of the wave vector.
		$\rmidx\xi{RI}$ ranges from $-1$ to $1$;
		both, $\rmidx\xi{R}$ and $\rmidx\xi{I}$, range from $-1$ to $1$ for even $n$ and from $-\IU$ to $\IU$ for odd $n$, respectively.
		Parameter values are
		$\timestep = \num{1.e-2}$,
		$\paramBl = \num{3.}$,
		$\paramBx = \num{3.5}$,
		$\paramD = \num{.1}$,
		$\paramE = \num{.1}$,
		$\paramF = \num{1.}$,
		$\modorifunc\dep{\nabla^2} = -\nabla^2$, and
		$n = 4$.
	}
	\label{F:MatrixElements}
\end{figure}

In \fref{F:MatrixElements} we plot the matrix elements of \eref{E:MatrixElements} as functions of $\veck$.
All of the matrix elements have poles at certain values of $\veck$.
These poles derive from a root of the determinant of $1 - \timestep L$, which enters into \eref{E:MatrixElements} as denominator.
The position of the poles depends on the physical parameters of the free energy functional
and also on the timestep $\timestep$ of Euler integration.
For sufficiently small $\timestep$, the poles fall outside the range of numerically accessible wave vectors.
Hence $\timestep$ can be chosen small enough to avoid numerical instability caused by the poles themselves.

Yet a different cause of instability remains, independent of $\timestep$:
In certain directions in reciprocal space the diagonal elements 
that correspond to the curves labeled by \circled{5} in \fref{F:MatrixElements}
exceed 1 for all wave vectors up to the pole.
These matrix elements let the high-frequency modes grow,
while the terms in $\rmidx\freeen{dir}$ of higher order in $\den$ and $\ori$ either overcompensate 
or fail to compensate for this growth.

To guarantee stability we want 
%% modified E-term for stability
the respective matrix elements to fall below 1, \ie
\begin{equation}
\label{E:relation}
	1 \geq \frac{a}{ad-\cnormsq{c_R}-\cnormsq{c_I}}
	\quad\Leftrightarrow\quad
	\paramF^2 \leq \frac{a \left( d-1 \right)}{\timestep^2 k^{2n}} .
\end{equation}
This relation can be used to compute the maximum $\paramF$, for which the calculations will be stable,
given the timestep $\timestep$, the maximum length of contributing wave vectors $k$, 
and the parameters of the free energy functional.
In addition, the relation \eref{E:relation} permits us to pinpoint a condition on the asymptotics of the $\paramE$-term,
necessary for numerical stability at arbitrary non-zero $\paramF$.
Let $\modorifunc\dep{-k^2}$ be of order $\LandauO{k^m}$ for large wave vectors $k \gg 1$.
Then \eref{E:relation} is expressed as
\begin{equation}
	\paramF^2 \leq \frac4{\timestep^2} \paramBx \paramE \, k^{4-2n} \modorifunc\dep{-k^2} + \LandauO{k^{m-2n+3}} .
	\label{E:Stability}
\end{equation}
In order to ensure that this relation holds for a finite $\paramF$ and arbitrarily large $k$,
$m \geq 2n - 4$ is required.
Conversely the function in the $\paramE$-term has to behave asymptotically like
\begin{equation}
	\modorifunc\dep{-k^2} \propto k^{2n-4} \qquad\text{for~} k \gg 1
\end{equation}
or like a term of higher order in $k$.
Only then the right hand side expression of \eref{E:Stability} does not tend to zero for large $k$.
%The same result is found, when the pseudodynamical equation for $\den$ is assumed to include a $\nabla^2$, instead of the Lagrange multiplier.

Note that qualitatively the order of the term $\modorifunc\dep{-k^2}$ seems not to matter as long as the integration is stable.
For example, for liquid crystals ($n=2$) we have found the same phases as in \cite{AchimPRE2011} 
even if we use $\modorifunc\dep{-k^2} = k^4$ instead of $\modorifunc\dep{-k^2} = k^2$.

\subsection{Overview of the phase behavior}

Due to the number of parameters and the rich zoo of complex phases that can occur, 
systematic studies of the whole parameter space are difficult. 
Our goal in this work is not to present complete phase diagrams.
In this and the following subsections we want to present typical phases 
that can occur in systems for various rotational symmetries.
Concerning the parameters we have focused on the regions in parameter space where non-trivial phases are expected 
as discussed 
\eg in the work by Achim \etal on \fold{2} particles \cite{AchimPRE2011}.

\begin{figure}
	\centering
	\includegraphics{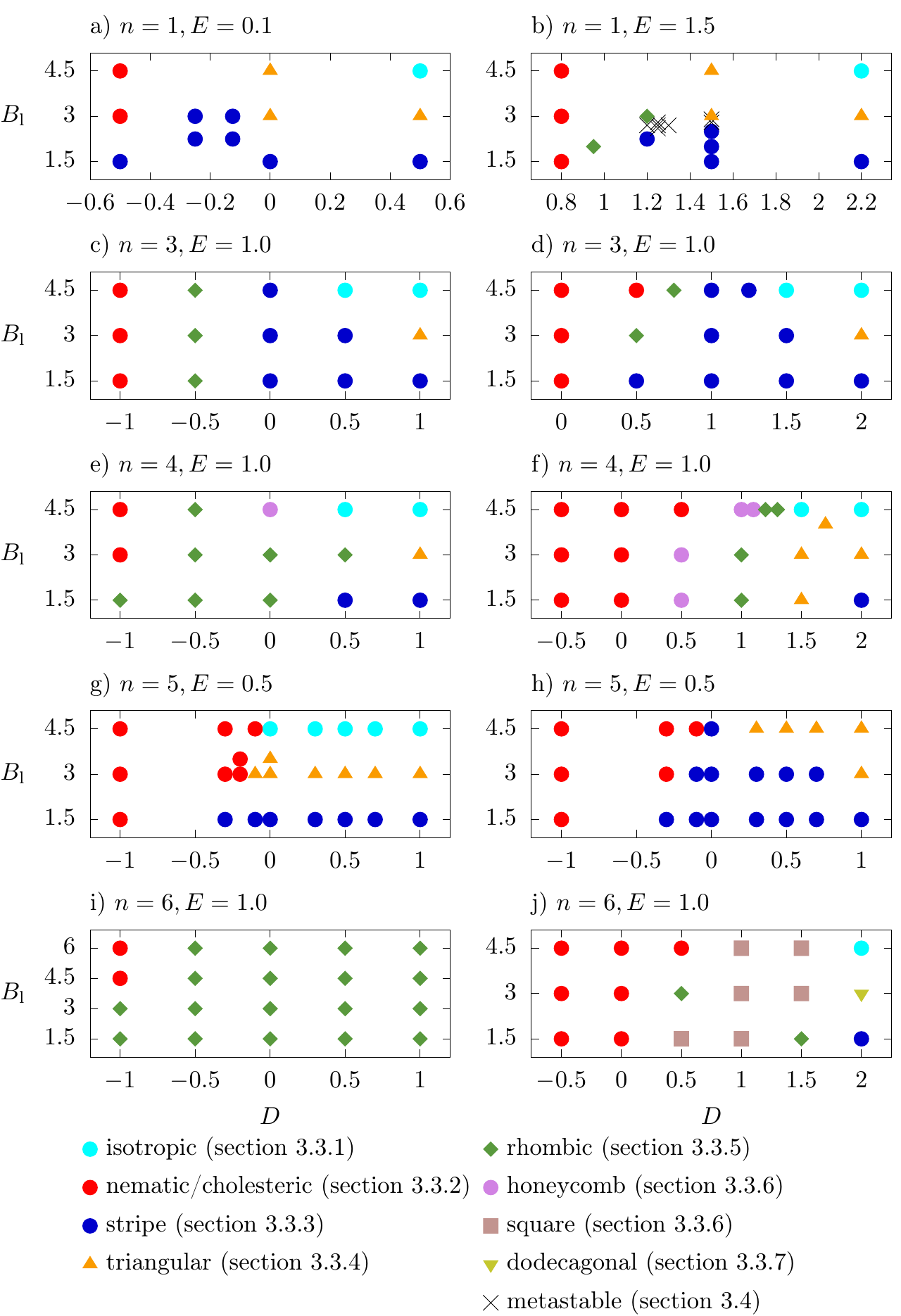}
	\caption{
		Left column: unmodulated,
		right column: modulated alignment.
		$\paramF = 0$ gives only isotropic, triangular, stripe with vanishing $\ori$, and nematic or cholesteric phase.
		We choose standard parameters $\paramBx = \num{3.5}, \paramF = \num{1.}$ and
		$\paramE = \num{.1}$ for unmodulated alignment and $\paramE = \num{1.}$ for modulated alignment, respectively.
		We have tried combinations of these values and $\paramBx = \num{1.}, \paramE = \num{3.5}, \paramF = \num{3.}$, without finding different phases.
	}
	\label{F:PhaseDiagrams}
\end{figure}

In \fref{F:PhaseDiagrams} we show an overview of the phases that we have found in systems with $1\leq n\leq 6$ 
both, for alignment between neighboring particles (left column in \fref{F:PhaseDiagrams}) 
as well as modulated alignments (right column). 
While the figures shown in \fref{F:PhaseDiagrams} not necessarily are complete phase diagrams 
as we cannot be sure whether there are additional intermediate phases (see also discussion in \sref{S:Metastable}),
the figures demonstrate the typical phases 
that occur in the small and large limits of $\paramD$ for almost all rotational symmetries
namely a nematic or cholesteric phase for small $\paramD$ and usually stripe, triangular, and isotropic phases for large $\paramD$.
Furthermore, in the overviews some non-trivial phases that occur in between are shown. 
Details of the phases are discussed in the following subsections as denoted in the legend of \fref{F:PhaseDiagrams}.

\subsection{Discussion of stable phases}

In the following we discuss the stable phases that we observe in detail.

Note that in all snapshots depicting the phases the color denotes the density-like field $\den\dep{\vecx}$.
The orientation field $\ori\dep{\vecx}$ is superimposed over $\den$ in the form of markers, 
which show the symmetry and direction. Furthermore, the size of the circle at the markers denotes magnitude $\cnorm{\ori}$.
Note that the fields $\ori$ and $\den$; are determined with the same spacial resolution. 
However, we plot fewer orientation markers to keep the plots legible.

\subsubsection{Isotropic phase}

For large $\paramD$ and large $\paramBl$ an isotropic phase is observed. In the isotropic phase the density is homogeneous
and the magnitude $\cnorm{\ori}$ of the orientation field vanishes, i.e., there is no preferred orientation of the particles.

\subsubsection{Nematic and cholesteric phases}

\begin{figure}
	\centering
	\includegraphics{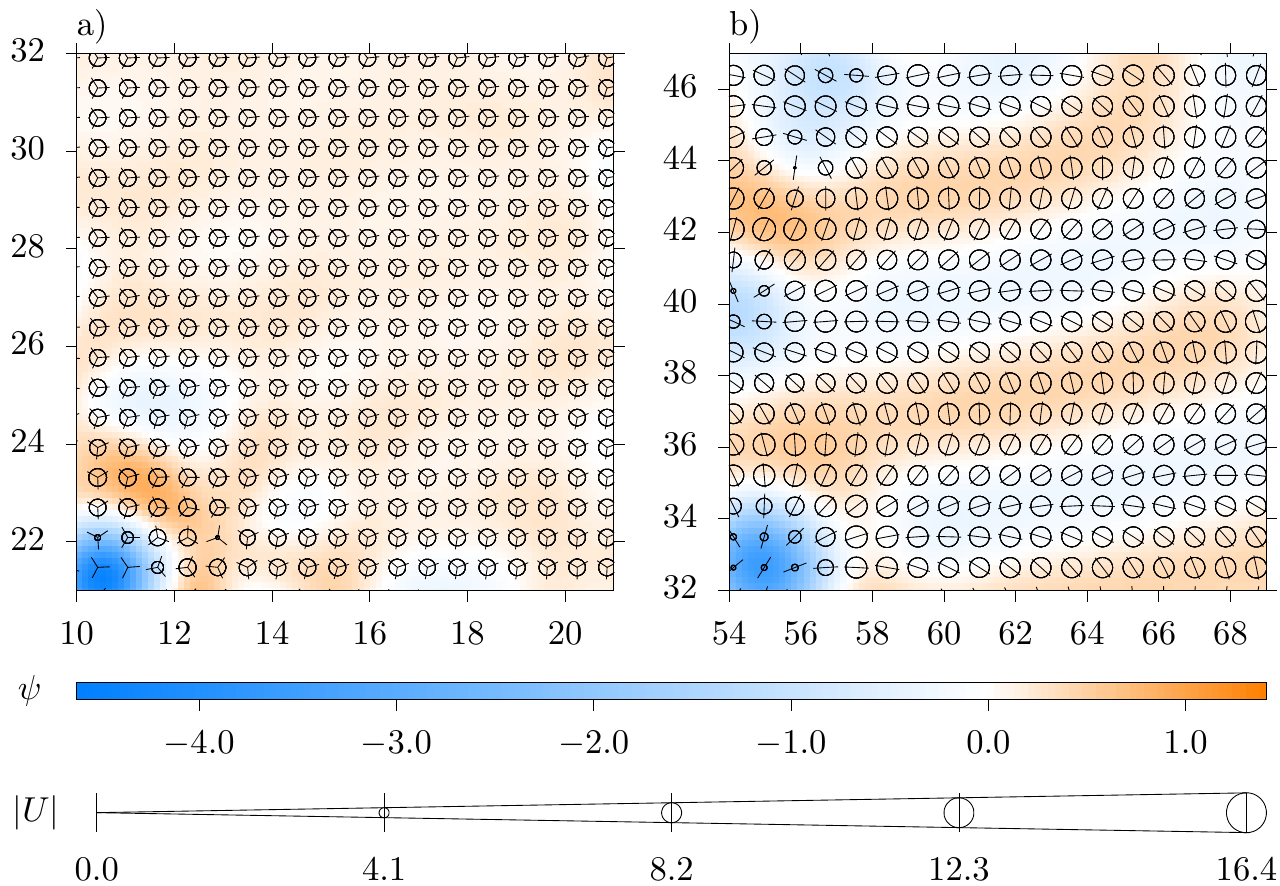}
	\caption{
		Phases that occur for small $\paramD$, where the orientational field is strong. 
		(a) Nematic phase for unmodulated alignment and \allparam{3}{3.}{3.5}{-1.}{.1}{1.}.
		(b) Cholesteric phase for modulated alignment and \allparam{2}{1.5}{3.5}{-1.}{.4}{1.}.
		In equilibrium the density in the nematic phase is constant 
		while the cholesteric phase possesses stripe-like density modulations.
		Large variations in density only appear in case of topological defects in the director field as depicted here
		in the lower left parts of the figures.
	}
	\label{F:Cholesteric}
\end{figure}

For small $\paramD$ structures with strong orientational order occur.
In case neighboring particles prefer to align \ie for unmodulated interactions a nematic phase is found
as depicted in \fref{F:Cholesteric}(a). 
If neighboring particle prefer opposite orientations \ie for modulated alignment interactions we observe a phase 
where a strong orientation changes continuously in a wave-like pattern. 
We call this structure a cholesteric phase.

While in an ideal nematic phase all particles possess the same orientation,
in the cholesteric phase particles with similar orientation occur along stripes. 
Furthermore, the density is constant in the nematic phase 
but there is a stripe-like weak density modulation in the cholesteric phase.

In both phases, when defects are present, they tend to cluster in motifs that occur in close-by phases.

\subsubsection{Stripe phases}

%even $n$: strong opposing ($\pi/n$) orientation on the maxima and minima; vanishing on the flanks
%odd $n$: strong opposing ($\pi/n$) orientation on the flanks; vanishing on the maxima and minima
\begin{figure}[h]
	\centering
	\includegraphics{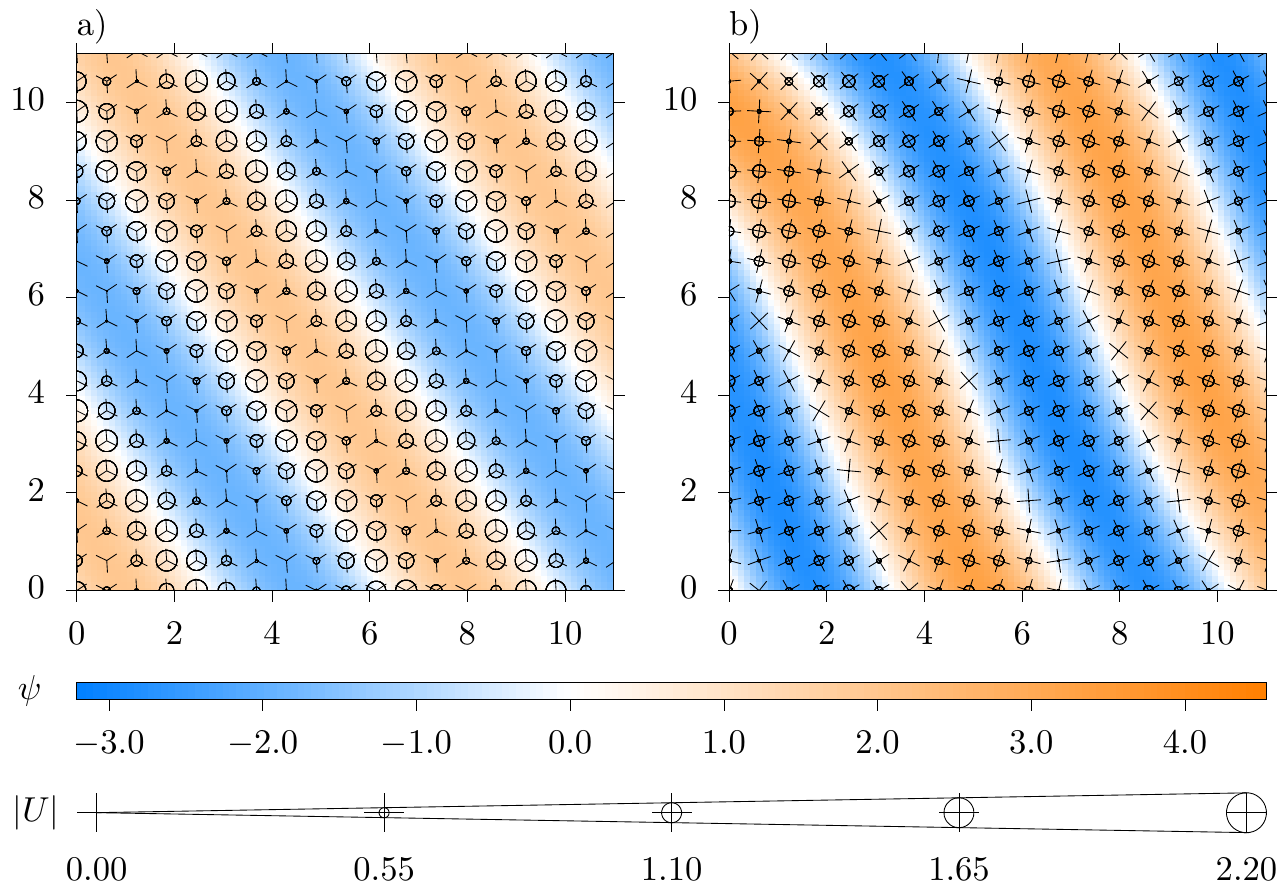}
	\caption{
		Typical stripe phases as they occur \eg for
		(a) unmodulated alignment and \allparam{3}{3.}{3.5}{.5}{.1}{1.} and
		(b) modulated alignment and \allparam{4}{1.5}{3.5}{2.}{1.}{1.}.
		The orientation fields of the stripe phases depend only on $n$ and are independent of (un)modulated alignment.
	}
	\label{F:Stripe}
\end{figure}

For small $\paramBl$ and usually larger $\paramD$ stripe phases with a strongly modulated density occur. 
The stripe phases of modulated and unmodulated alignment are identical.
For all odd $n$ we observe the strongest modulations to appear where the density gradient is maximal 
as depicted in \fref{F:Stripe}(a). 
In contrast, for all even $n$, \fref{F:Stripe}(b), the orientational strength is maximal 
in the density maximum as well as in opposite direction (rotated by $\pi/n$) in the density minimum but vanishes at the flanks of the stripes.

\subsubsection{Triangular phases}

\begin{figure}
	\centering
	\includegraphics{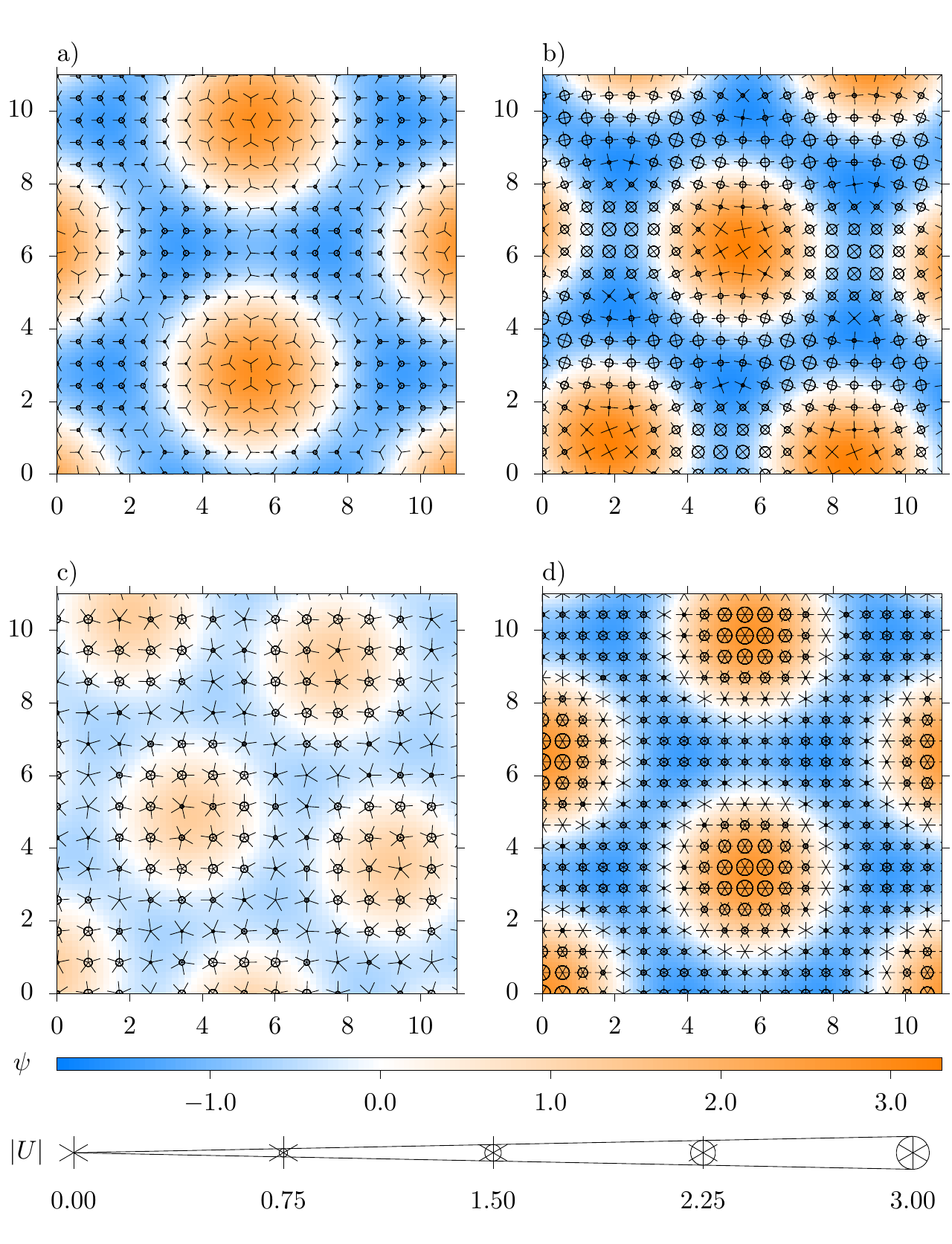}
	\caption{
		Triangular phases for
		(a) unmodulated alignment and \allparam{3}{3.}{3.5}{1.}{.1}{1.},
		(b) modulated alignment and \allparam{4}{3.}{3.5}{1.5}{1.}{1.},
		(c) modulated alignment and \allparam{5}{4.5}{3.5}{1.}{.5}{1.}, and
		(d) modulated alignment and \allparam{6}{3.}{3.5}{1.}{1.}{1.}.
		In (d) $\cnorm{\ori}$ is twice as large as plotted and the phase is metastable.
	}
	\label{F:Triangular}
\end{figure}

The triangular phases occur at larger $\paramBl$ than the stripe phases.
Similar to the stripe phases, the orientation fields of the triangular phases depend whether $n$ is even or odd;
yet with exceptions when $n$ matches a symmetry of the triangular lattice.
Typically for even $n$ the orientation is strong between neighboring density peaks, see \Fref{F:Triangular}(b),
while for odd $n$ each density peak is surrounded by a ring of strong orientation as shown in \Fref{F:Triangular}(c).
In both cases the orientation vanishes at the density maxima, marking the triangular phases as plastic crystals.
Exceptions of this general rule are $n = 3$ and $n = 6$:
\Fref{F:Triangular}(a) shows the triangular phase of $n = 3$, exhibiting strong orientation not between pairs, 
but between triplets of neighboring density peaks.
\Fref{F:Triangular}(d) presents the triangular phase of $n = 6$, which is an oriented crystal.
However, it only occurs as metastable phase as the dodecagonal quasicrystal is stable for these parameters.

\subsubsection{Rhombic phases}

\begin{figure}
	\centering
	\includegraphics{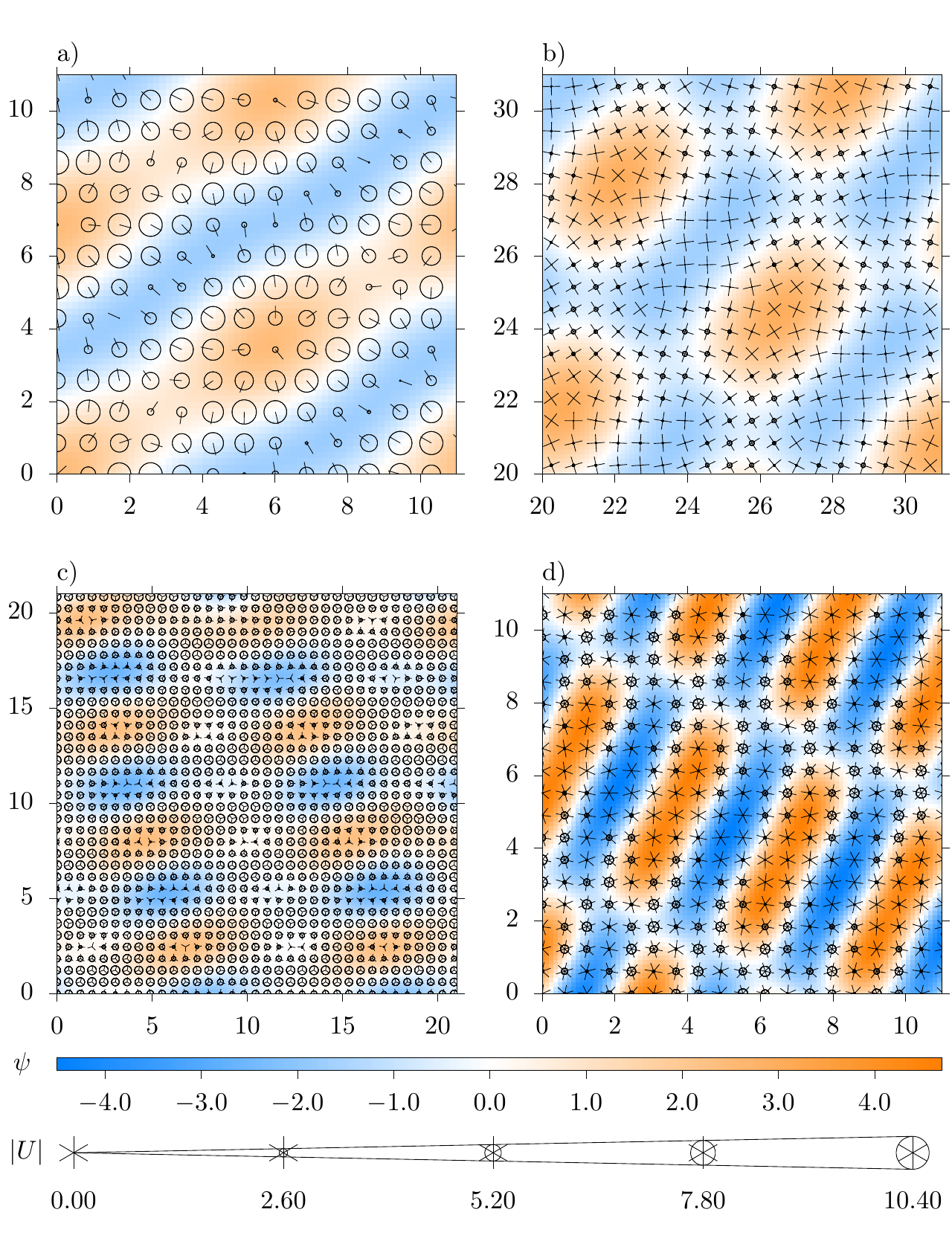}
	\caption{
		Rhombic phases for
		(a) modulated alignment and \allparam{1}{2.}{3.5}{.95}{1.5}{1.},
		(b) unmodulated alignment and \allparam{4}{3.}{3.5}{.5}{1.}{1.},
		(c) modulated alignment and \allparam{5}{4.5}{3.5}{.5}{1.}{1.}, and
		(d) unmodulated alignment and \allparam{6}{6.}{3.5}{.5}{.1}{1.}.
	}
	\label{F:Rhombic}
\end{figure}

\Fref{F:Rhombic} shows several rhombic phases for different $n$.
The rhombic phases often appear at intermediate $\paramD$ between the nematic or cholesteric phase and other crystalline phases.
In some cases the angles that occur might not be uniquely given:
For modulated alignment next to the cholesteric phase, 
the rhombic phase often can be stretched or compressed at almost no energy cost,
because there is a flat segment in the piecewise definition of $\modorifunc\dep{-k^2}$ in \eref{E:ModOri}.

\subsubsection{Honeycomb and Square phases}

\begin{figure}[h]
	\centering
	\includegraphics{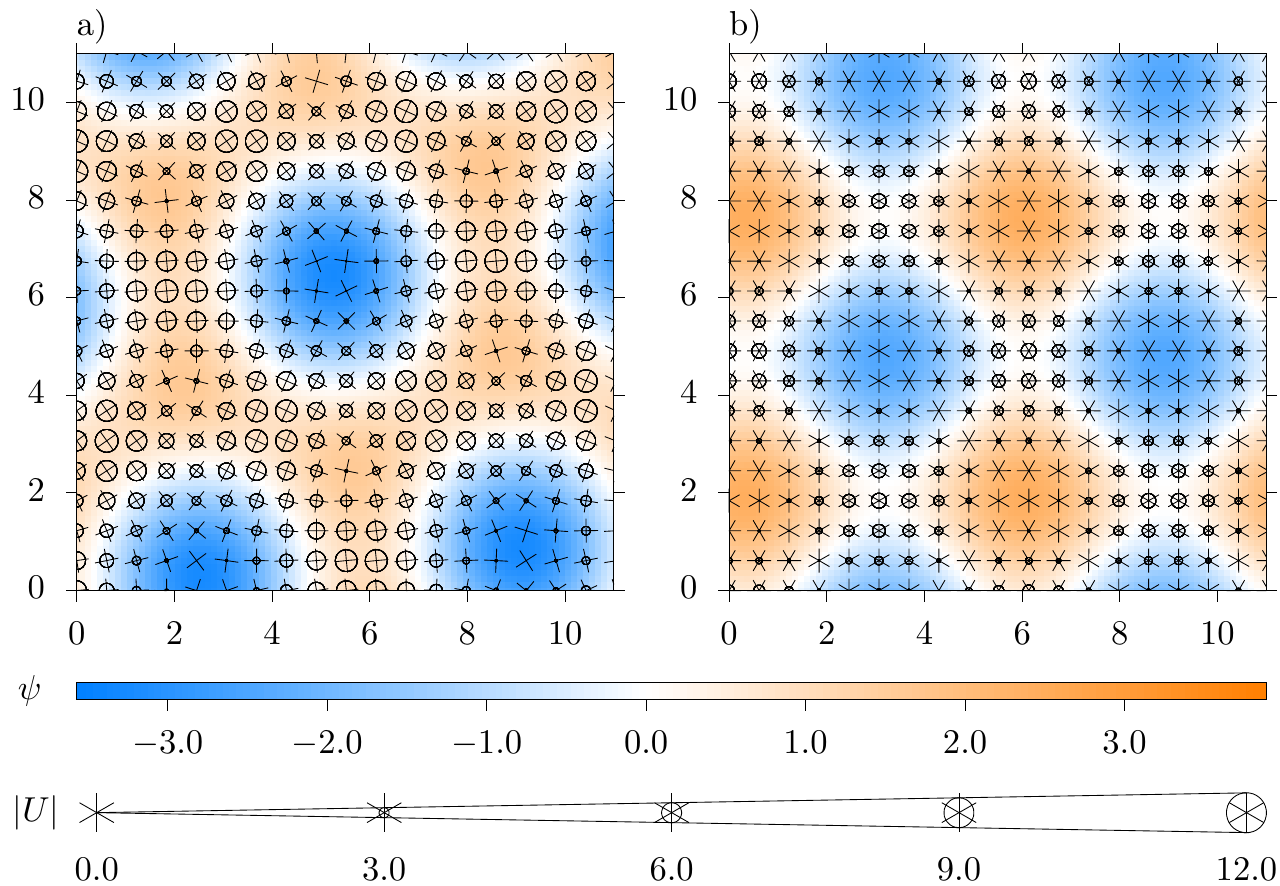}
	\caption{
		(a) Honeycomb phase for modulated alignment and \allparam{4}{3.}{3.5}{.5}{1.}{1.} and
		(b) square phase for modulated alignment and \allparam{6}{4.5}{3.5}{1.}{1.}{1.}.
	}
	\label{F:HoneycombSquare}
\end{figure}

The honeycomb and square phases, depicted in \fref{F:HoneycombSquare}, 
add to the examples of plastic crystalline phases which appear in between the ubiquitous phases.
Their topology is the same as reported for \fold{2} particles in \cite{AchimPRE2011}.

\subsubsection{Dodecagonal quasicrystalline phase}
\label{S:dodecagonal}
\begin{figure}
	\centering
	\includegraphics{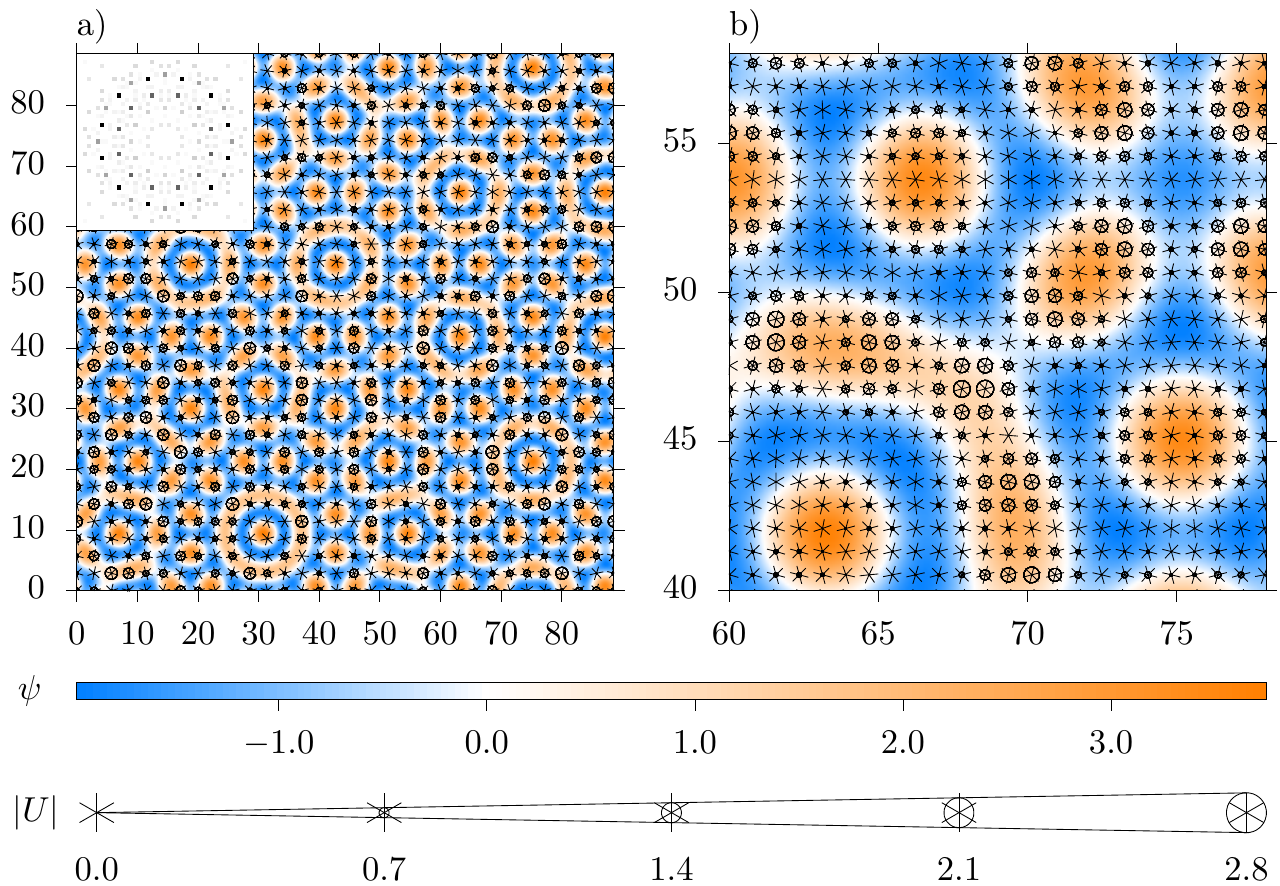}
	\caption{
		Dodecagonal quasicrystal for modulated alignment and \allparam{6}{3.}{3.5}{2.}{1.}{1.}.
		The inset shows the Fourier transform of $\den$ and 
		(b) is a zoom-in on the configuration from (a).
	}
	\label{F:Dodecagonal}
\end{figure}

We also find a stable quasicrystal, \ie a structure with long-ranged order but without translational symmetry. 
The observed quasicrystal is shown in \fref{F:Dodecagonal}. 
It possesses \fold{12} rotational symmetry and occurs for a system with modulated alignments 
and particles with \fold{6} rotational symmetry. 
From a zoom-in to the quasicrystalline structure as shown in \fref{F:Dodecagonal}(b) 
one recognizes that along a ring around a local symmetry center there are 12 regions with strong orientation.
The orientation in neighboring regions is rotated by $\pi/6$ as expected due to the modulation of the alignment interaction.
Probably these regions are the features that stabilize the quasicrystals.

As typical to quasicrystals, the Fourier transform of $\den$ exhibits not only the 12 main peaks 
but numerous quite strong satellite peaks as well.
Therefore, the anisotropic interaction can indeed lead to the stabilization of lengthscales 
that differ from the length that is preferred by the free energy functional.

In principle a similar structure might occur for particles with \fold{12} rotational symmetry and unmodulated alignment interactions.
However, we have not found such a quasicrystal. Furthermore, we have not observed any stable quasicrystals with other rotational symmetry.

\subsection{Defects and metastable states}
\label{S:Metastable}
\begin{figure}
	\centering
	\includegraphics{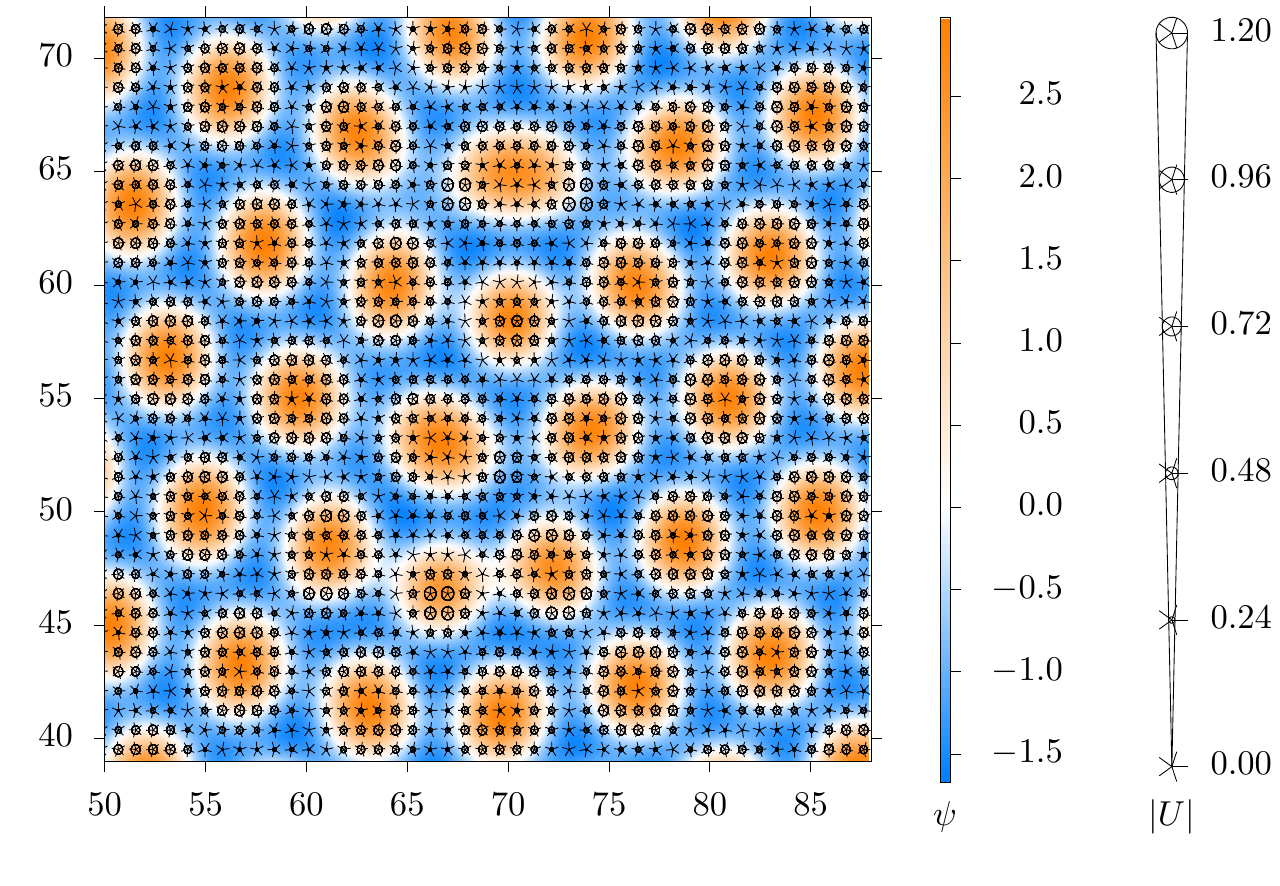}
	\caption{
		Defects along a grain boundary of the triangular phase of \fold{5} particles
		for unmodulated alignment and $\paramBl = \num{3.}, \paramD = \num{.7}, \paramE = \num{.5}$. %\allparam{5}{3.}{3.5}{.7}{.5}{1.}.
		The stable phase for these parameters is a triangular crystal.
		The orientation field is large on the density peaks with 5 neighbors.
	}
	\label{F:FiveFoldDefects}
\end{figure}

Although we do not observe any stable quasicrystals with \fold{5} rotational symmetry,
it seems that \fold{5} symmetry prominently occurs in defects along grain boundaries as shown in \fref{F:FiveFoldDefects}.

\begin{figure}
	\centering
	\includegraphics{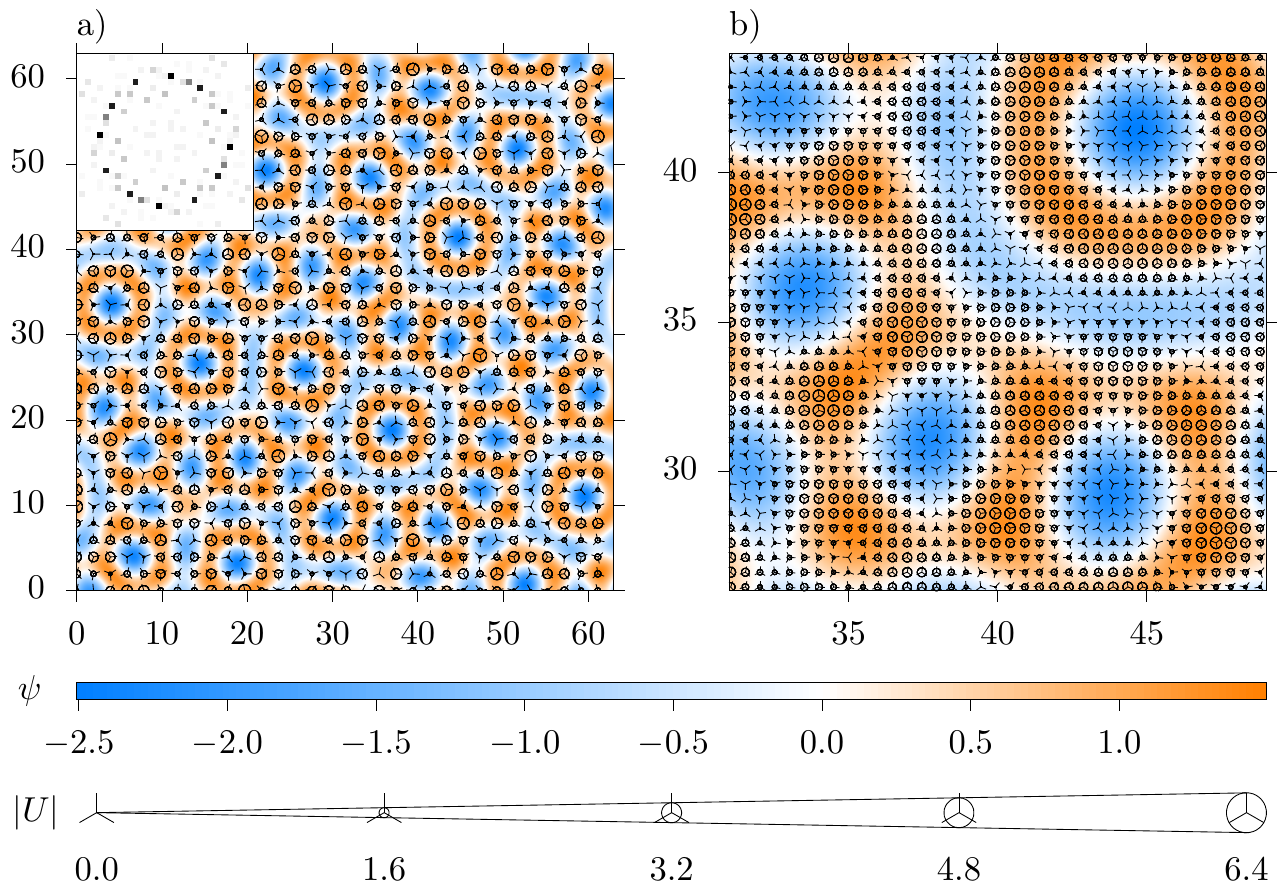}
	\caption{
		Metastable dodecagonal quasicrystal for modulated alignment and \allparam{3}{4.5}{3.5}{1.}{1.}{1.}.
		Note that the stable phase for these parameters is the stripe phase.
		The inset shows the Fourier transform of $\den$ and 
		(b) is a zoom-in on the configuration from (a).
	}
	\label{F:MetastableDodecagonal}
\end{figure}

\begin{figure}
	\centering
	\includegraphics{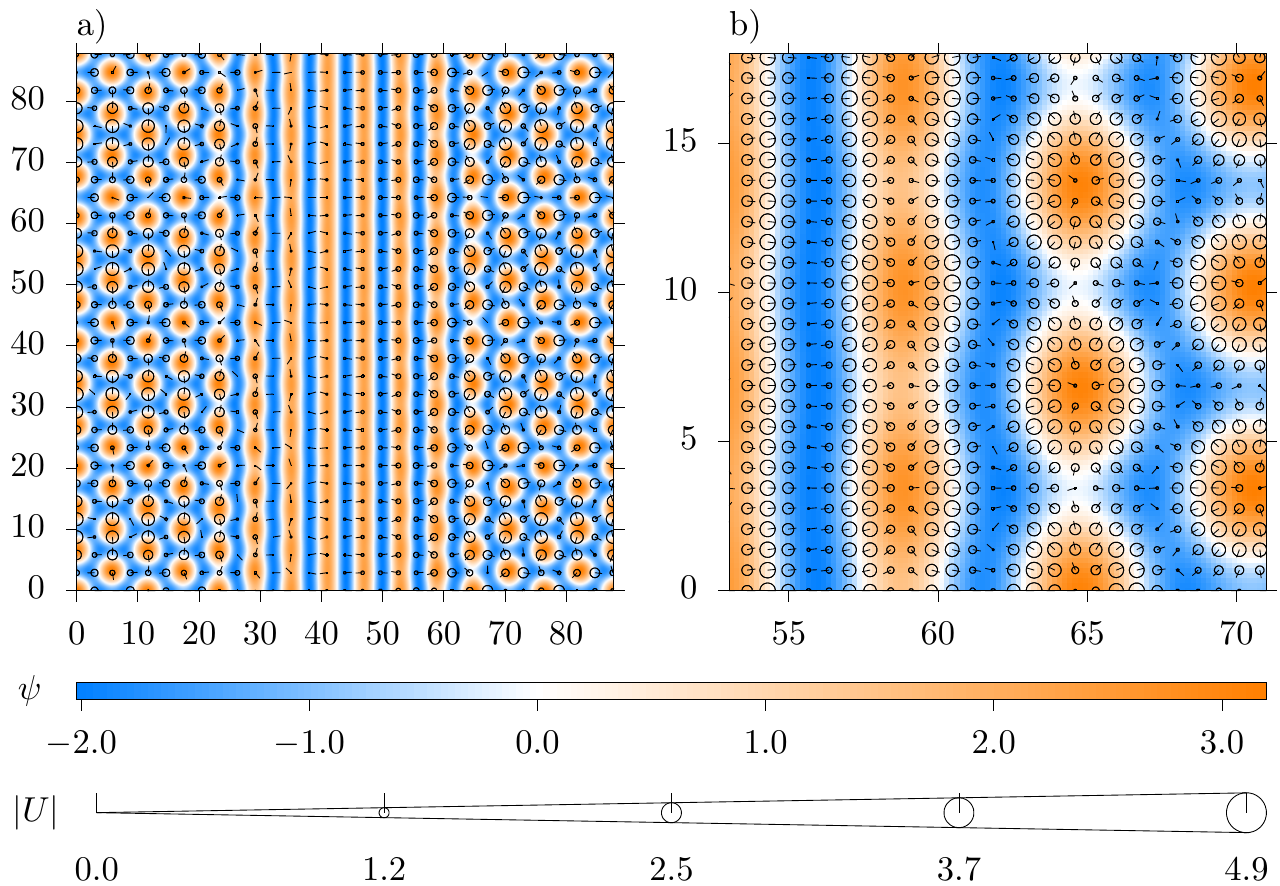}
	\caption{
		Mixture of triangular, stripe, and possibly rhombic phase for modulated alignment and \allparam{1}{2.9}{3.5}{1.5}{1.5}{1.}.
		(b) is a zoom-in on the configuration from (a).
	}
	\label{F:MetastableTriangular}
\end{figure}

We also observe metastable quasicrystals like the one shown in \fref{F:MetastableDodecagonal} that are different from the dodecagonal structure reported in \sref{S:dodecagonal}.

The figures shown in \fref{F:PhaseDiagrams} are not necessarily complete phase diagrams. 
In some cases it is hard to figure out whether a resulting structure is stable or metastable.
For example in \fref{F:PhaseDiagrams}(b) some points in parameter space are marked with crosses. 
For these parameters we observe structures that seem to be mixtures of triangular, stripe, and possibly rhombic phase as shown
in \fref{F:MetastableTriangular}.
Note that the corresponding pure phases possess a higher free energy than the mixture that we observe.
Probably in these points the true equilibrium phase has not been found yet.
%%%%%%%%%%%%%%%%%%

%\input{conclusions.ltx}
\section{Conclusions}
\label{S:Conclusions}

We have introduced a new \PFC model for particles with $n$-fold rotational symmetry in two dimensions as it occurs, 
\eg for patchy particles with symmetrically placed patches. 
Both the cases of attractive as well as repulsive patches have been considered leading 
either to alternating or the same orientation of neighboring clusters.

The \PFC model is used to determine the phases that are stable for various rotational symmetries. 
We usually observe nematic or cholesteric phases in case a non-vanishing orientational order is preferred by the free energy.
In the opposite limit we find stripe, triangular, or isotropic phases depending on how strong density modulations are supported.
In between these phases, complex orderings with honeycomb, square, or rhombic symmetry occur. 
We even find a quasicrystalline phase with dodecagonal rotational symmetry.
Therefore it is demonstrated that quasicrystals can be stabilized by interactions 
that only possess one length scale if in addition special binding angles are preferred.
Note that we do not find the type of quasicrystals 
that has been reported to occur for patchy particles in simulations 
\cite{vanderlinden2012,reinhardt2013,GemeinhardtEPJE2018,GemeinhardtEPL2019}. 
As we will discuss in the last paragraph this is probably due to lack of a hard core in our approach.

Since a lot of complex phases occur in our system, finding the global minimum might be hard 
and usually there are a lot of metastable states with interesting structures. 
At the end of section 3 we shortly comment on some examples including a metastable quasicrystal. 
However, the metastable states deserve more detailed analyses in future works. 
For example, the domains that meet at grain boundaries prefer orientations 
that depend on the rotational symmetry and the orientational interactions.
As a consequence the grain boundaries probably differ from the boundaries that are observed for isotropic particles.
By adding noise the coarsening processes in systems with such grain boundaries can be explored, which we leave for future works.

The particles that we have in mind in this work do not possess a hard core. 
Therefore, the phases described here correspond to cluster crystals similar as in \cite{Sciortino2004,Mladek2006,Lenz2012,Sciortino2013,Stiakakis2021, Barkan2014}.
In contrast, in many particulate systems a hard core prevents large overlaps of particles. 
Furthermore, a hard core can support the formation of some complex phases like quasicrystals \cite{dotera2014,Fayen2020,Fayen2022}. 
Note that the quasicrystalline phases that have been observed in computer simulations 
of patchy colloids have been found in systems where the particles can hardly overlap 
\cite{vanderlinden2012,reinhardt2013,GemeinhardtEPJE2018,GemeinhardtEPL2019}.
As a consequence, in future we want to study mean field approaches with similar couplings between an orientational field and the density-like field as in this article.
However, to model a hard core as well, the functional dependence on the density-like field has to be changed. 
A suitable way to describe the hard core is given by the so-called fundamental measure theory \cite{Rosenfeld1989,HansenGoos2006,Roth2010},
which can be formulated in two dimensions \cite{Roth2012} 
and which is known to lead to complex phases in case of the competition with an incommensurate substrate \cite{Neuhaus2013,Neuhaus2013b,Neuhaus2014}.
%%%%%%%%%%%%%%%%%%

\section*{Data availability statement}
The data that support the findings of this study are available upon reasonable request from the authors.

\ack	% -nowledgments
This work is supported by the Deutsche Forschungsgemeinschaft (Grant SCHM 2657/4-1).

\section*{References}
\bibliography{literature}
\bibliographystyle{iopart-num}

\end{document}